\documentstyle[12pt]{article}
\newcommand{\eqref}[1]{(\ref{#1})}   
\newcommand{\naive}{na\"{\i}ve}
\newcommand{\eff}{{\hbox{\footnotesize\it eff}\,}}

\newcommand{\GeV}{\hbox{$\hbox{GeV}^2$}}
\newcommand{\Du}{\Delta u}
\newcommand{\Dd}{\Delta d}
\newcommand{\Ds}{\Delta s}
\newcommand{\xQ}{(x,Q^2)}

\newcommand{\gIp}{g_1^p(x,Q^2)}
\newcommand{\gIn}{g_1^n(x,Q^2)}
\newcommand{\gId}{g_1^d(x,Q^2)}

\newcommand{\AIp}{A_1^p(x,Q^2)}
\newcommand{\AIn}{A_1^n(x,Q^2)}

\newcommand{\GIp}{\Gamma_1^p(Q^2)}
\newcommand{\GIn}{\Gamma_1^n(Q^2)}
\newcommand{\GId}{\Gamma_1^d(Q^2)}
\newcommand{\GIpn}{\Gamma_1^{p-n}(Q^2)}

\newcommand{\pr}{\paragraph{}}
\newcommand{\beq}{\begin{equation}}
\newcommand{\eeq}{\end{equation}}
\newcommand{\bea}{\begin{eqnarray}}

\newcommand{\eea}{\end{eqnarray}}

\begin{document}

\begin{titlepage}
\begin{flushright}
CERN-TH-6898/93\\
TAUP-2052-93\\
hep-ph/9305306
\end{flushright}
\begin{centering}
\vspace{.1in}
{\large {\bf Analysis of Data on Polarized Lepton-Nucleon
Scattering}} \\
\vspace{.4in}
{\bf John Ellis}\\
\vspace{.05in}
Theory Division, CERN, CH-1211, Geneva 23, Switzerland. \\
e-mail: johne@cernvm.cern.ch \\
\vspace{0.5cm}
and \\
\vspace{0.5cm}

\vspace{.05in}
{\bf Marek Karliner}\\

\vspace{.05in}
School of Physics and Astronomy
\\ Raymond and Beverly Sackler Faculty of Exact Sciences
\\ Tel-Aviv University, 69978 Tel-Aviv, Israel
\\ e-mail: marek@vm.tau.ac.il
\\
\vspace{.03in}
\vspace{.1in}
{\bf Abstract} \\
\vspace{.05in}
\end{centering}
{\small
We re-analyze data on deep inelastic polarized lepton-nucleon
scattering, with particular attention to testing the Bjorken sum
rule and estimating the quark contributions to the nucleon spin.
Since only structure function data at fixed $Q^2$ can be used to test
sum rules, we use E142 asymmetry measurements and unpolarized structure
function data to extract $g_1^n$ at fixed $Q^2$ = 2 GeV$^2$. When
higher-twist effects, which are important at low $Q^2$, are included,
both the
E142 and SMC data are compatible with the Bjorken sum
rule within one standard deviation. Assuming validity of the Bjorken
sum rule, we estimate the quark contributions to the nucleon spin,
finding that their total net contribution is small, with the strange
quark contribution non-zero and negative. The quark spin content of
the nucleon spin is in agreement with Skyrme model.}

\paragraph{}
\par
\vspace{1.0in}
\begin{flushleft}
CERN-TH-6898/93\\
May 1993 \\
\end{flushleft}

\end{titlepage}
\newpage

\section{Introduction}

It was the Bjorken sum rule \cite{BJ},\cite{Kodaira}
for polarized lepton-nucleon scattering,
derived initially \cite{BJ}
 from quark current algebra at short distances, that
led to the first proposal that deep inelastic structure functions
should scale. However, although approximate scaling has been observed
in deep inelastic scattering and quark model relations for
structure functions hold in general, with deviations which are
understood quantitatively in terms of perturbative QCD and
asymptotic freedom at large momentum transfers,
the Bjorken sum rule was never actually tested
during the 25 years following its discovery. Data on polarized
lepton-proton scattering have been available for about
15 years \cite{oldSLACa}-\cite{EMCNP}, but
data on polarized lepton-neutron scattering have become available
only very recently \cite{SMC},\cite{E142}.
The stimulus for performing these experiments,
finally, was provided by the results of the
EMC experiment \cite{EMCPL},\cite{EMCNP} on
polarized muon-proton scattering, which differed from predictions
based on \naive\ constituent quark model ideas \cite{EJ},
but were consistent with
perturbative QCD. Polarized lepton-neutron scattering would also have
to differ from \naive\ constituent quark model ideas if the Bjorken
sum rule were to be valid. If it were not valid, the whole edifice of
perturbative QCD and asymptotic freedom would collapse \cite{PRthat}.
\pr
One of the two results on polarized lepton-neutron scattering
announced recently, the one  by the SMC \cite{SMC},
does indeed find a deviation
from \naive\ constituent quark model ideas, and, when combined with
the EMC proton data, verifies the Bjorken sum rule within admittedly
large errors. However, the other polarized neutron experiment
by the E142 collaboration \cite{E142}, which has much
smaller statistical errors, is claimed, when combined with the EMC data,
to disagree with the Bjorken sum rule by two standard deviations. This
result therefore presents a puzzle whose resolution apparently needs
affirmative answers to one or more of the following questions: are
the EMC data wrong? are the E142 data wrong? is the Bjorken sum
rule wrong?
\pr
We take the point of view in this paper that one must be prudent
before jumping to any such dramatic conclusion, especially before
concluding that the Bjorken sum rule is wrong. We recall that in
every polarization experiment the
data have been taken in different ranges of $Q^2$, which
varied from bin to bin in the Bjorken scaling variable $x$, with
average $Q^2$ values that are known in the case of the SMC and
EMC experiments,
but have not been published by the E142 collaboration. The EMC and
the SMC have used the fact, observed also in other experiments, that
the asymmetry $A_1(x)$ has no detectable dependence on $Q^2$ in the
range studied, to estimate the polarized structure functions at a
fixed value of $Q^2$. Whereas the EMC used their own measurements of
the unpolarized structure function $F_2(x,Q^2)$ and a perturbative
QCD model for the ratio $R(x,Q^2)$ of longitudinal to transverse
virtual photon cross-sections
to calculate $g_1(x,Q^2)$ for $Q^2$ fixed at
$10.7$ \GeV, the SMC used the more recent
NMC parametrization \cite{NMC}
of the unpolarized structure function $F_2(x,Q^2)$
and a SLAC parametrization \cite{RLT} of the ratio $R(x,Q^2)$
to calculate $g_1(x,Q^2)$ for $Q^2$
fixed at $4.6$ \GeV. This procedure is {\em a priori} better
suited for testing sum rule predictions, which are formulated for
structure functions integrated over all $x$ values at fixed $Q^2$,
than would be an integral over $x$ at variable values of $Q^2$.
\pr
In this paper, we first check whether there is indeed any evidence
against the validity of the Bjorken sum rule. To do this, we start by
rescaling the EMC polarized proton data \cite{EMCNP}
to $Q^2$ at $4.6$ \GeV, for
comparison with the SMC \cite{SMC},
 and at $2$ \GeV, for comparison with E142 \cite{E142}
using the latest NMC parametrization \cite{NMC}
of the unpolarized structure
function and the latest SLAC parametrization of $R(x,Q^2)$ \cite{RLT}.
 For the
latter comparison, we use the measurements of $A_1^n$ reported by
E142 and the same NMC and SLAC parametrizations to extract $g_1^n(x,Q^2)$
for $Q^2$ fixed at $2$ \GeV. Next, we discuss the uncertainties in
the extrapolations of the E142 data to $x$ = $0$ and $1$. Since the
appropriate power behaviour at small $x$ is not tightly constrained
by other data or by theory, this introduces some extra uncertainty
beyond that discussed by E142. $A$ $priori$, the error in the SMC
extrapolation to $x$ = $0$ is $5$ times smaller, since they have
measurements down to $x$ = $0.006$, as opposed to the $0.03$ of E142.
Contrary to what is stated in ref.~\cite{E142},
the behaviour of the polarization asymmetry at large $x$ is not
determined by perturbative QCD without extra assumptions about the
non-perturbative structure of the nucleon, which also tends to
increase the uncertainty beyond that assumed by E142.
Next we show that, when proper attention is paid to the $Q^2$
dependence of the structure functions at large $x$, the SMC
data are largely compatible with a recently derived bound
\cite{PRbound}.
We then extract
the integrals of
\hbox{$g_1^p$ - $g_1^n$} at fixed $Q^2$ = $4.6$ \GeV\
by combining the rescaled EMC and SMC data, and at $2$ \GeV\ by
combining the rescaled EMC and E142 data, including realistic
allowances for the errors. Particularly in the low-$Q^2$ case of the
comparison of E142 and the EMC, higher-twist and mass corrections
must be taken into account. According to the indicative estimate in
ref.~\cite{BBK}
these corrections are more than $10$\% at $Q^2$ = $2$ \GeV\ for the
Bjorken sum rule.
\footnote{See however ref.~\cite{BI} for a recent discussion
of higher-twist effects in this context.}
The estimates of ref.~\cite{BBK} should therefore
be included in the theoretical prediction.
After including these subasymptotic effects, both
the E142 as well as the SMC data are in good agreement with the
Bjorken prediction. Encouraged by this agreement, we assume that
the Bjorken sum rule is indeed correct, and go on to extract the
contributions of the various quark flavours to the proton spin
using all the available polarized structure function data.
We
find
\beq
\Du + \Dd + \Ds = 0.22\,\pm\,0.10
\label{Dqformula}
\eeq
which is far from the \naive\ quark model. It is close to the prediction
of the Skyrme model for light quarks \cite{BEK}.
We conclude that a quantitative understanding of axial current matrix
elements is now emerging.
\pr
\section{Data at Fixed $Q^2$}
\pr
We first discuss the rescaling of the available polarized structure
function data to a fixed value of $Q^2$, which is a crucial step for
testing QCD sum rules, that are formulated at fixed $Q^2$. Experiments
measure directly the polarization asymmetry $A_1(x,Q^2)$, in terms of
which the polarized structure function $g_1(x,Q^2)$ is given by
\beq
g_1(x,Q^2) = {A_1(x,Q^2) F_2(x,Q^2) \over 2 \,
x \left[ 1+R(x,Q^2)\right] }
\label{gIdef}
\eeq
where $F_2(x,Q^2)$ is the conventional unpolarized structure
function, and $R(x,Q^2)$ is the ratio of virtual longitudinal
to transverse virtual photon cross
sections. As already mentioned, each experiment takes
data over a range of $Q^2$ that varies with $x$, and differs from
experiment to experiment:
$\langle Q^2\rangle$ = $2$, $4.6$, $10.7$ \GeV\ for E142, the
SMC and the EMC respectively. Each experiment reports that the
asymmetry that it measures exhibits no detectable variation with
$Q^2$ in any $x$ bin. It is therefore reasonable to use this
$Q^2$-independent value, together with the unpolarized $F_2(x,Q^2)$
and $R(x,Q^2)$ measured in other experiments, to estimate $g_1(x,Q^2)$
at some fixed $Q^2$, taken ideally to be the average
$\langle Q^2\rangle$ for
each experiment. However, to combine the experiments that have
different
$\langle Q^2\rangle$ to test the Bjorken
sum rule, it is necessary to estimate
$g_1(x,Q^2)$ for at least one non-ideal value of $Q^2$: the $Q^2$
variation of $g_1(x,Q^2)$ is substantial, even if that of $A_1(x,Q^2)$
is negligible.
\pr
Both the EMC \cite{EMCNP}
and the SMC \cite{SMC} have presented their data at fixed $Q^2$
as advocated above. However, the EMC used a parametrization of the
unpolarized data that has now been superseded by more recent results
from the NMC \cite{NMC}
and SLAC \cite{RLT}. The latter have been used by the SMC to
estimate their $g_1^d(x,Q^2)$, and we use them here to re-estimate
$g_1^p(x,Q^2)$. Figure~1 shows the
rescaled EMC $\gIp$ at fixed $Q^2$ = $2$, $4.6$ and $10.7$ \GeV,
using a smooth parametrization \cite{EMCNP} of the $A_1^p\,\,$
 EMC data.
The two former values of $Q^2$
are needed for combining with the E142 and
SMC data respectively: the $Q^2$-dependence of $g_1^p(x,Q^2)$ is
seen clearly, and the corresponding integrals
$\Gamma_1^p(Q^2)\equiv \int_0^1\gIp$ are
shown in Table 1.
\pr
\vbox{
\begin{centering}
{\bf Table 1}\\
\vskip 0.4cm
\begin{tabular}{||c|c||} \hline\hline
$ Q^2 (\GeV)  $ & $ \GIp $\\ \hline
$ 2.0         $ & $ 0.124 \,\pm\,0.013 \,\pm\,0.019 $ \\ \hline
$ 4.6         $ & $ 0.125 \,\pm\,0.013 \,\pm\,0.019 $ \\ \hline
$ 10.7        $ & $ 0.128 \,\pm\,0.013 \,\pm\,0.019 $ \\
\hline\hline
\end{tabular}\\
\end{centering}
\vskip 1cm
\centerline{
\vbox{\hsize=14cm
\footnotesize
\noindent
The integrals $\Gamma_1^p(Q^2)\equiv \int_0^1\gIp$
obtained from the EMC \cite{EMCNP} data on $\AIp$,
at $Q^2=2$ \GeV, combined with the NMC \cite{NMC} data
for $F_2^p\xQ$ and SLAC parametrization \cite{RLT} of
$R\xQ\equiv \sigma_L/\sigma_T$.
The first is the statistical and the second is the systematic
error.
} 
}
} 
\noindent
We assume here that the systematic error on $\Gamma_1^p$ is as quoted
by the EMC \cite{EMCNP}, even though the systematic error on the NMC
$F_2^p$ is smaller than the systematic error on the EMC $F_2^p$. If
there were no error at all on $F_2^p$ the systematic error in
$\Gamma_1^p$ would be reduced from $\pm0.019$ down to $\pm 0.018$.
The difference from the value quoted by the EMC
for $Q^2$ = $10.7$ \GeV\ is well within their quoted errors: we
will comment later on the $Q^2$-dependence of $\GIp$.
\pr
We have also applied the same procedure to the E142 data \cite{E142},
using their values of $A_1^n(x)$ and the NMC and SLAC
parametrizations of $F_2(x,Q^2)$ and $R(x,Q^2)$ respectively
to estimate $g_1^n(x,Q^2)$ at $Q^2$ = $2$ \GeV, their average value.
This procedure yields
\beq
\int_{0.03}^{0.6} d x g_1^n(x,Q^2{=}2\,\GeV) =
-0.022\,\pm\,0.006\,\hbox{(stat.)}\,\pm\,0.006\,\hbox{(syst.)}
\label{ourSLACbulk}
\eeq
to be compared with the estimate
of $-0.019$ $\pm\ 0.006$ (stat.) $\pm\ 0.006$ (syst.) given by E142.
\pr
Before estimating $\Gamma_1^n(Q^2=\GeV)$ on the basis of equation
\eqref{ourSLACbulk},
we first comment on the extrapolations of their data beyond $x$ = $0.6$
and below $x$ = $0.03$. Models for the polarization asymmetry that
combine a non-perturbative Ansatz for the neutron wave function with
perturbative QCD at large $Q^2$ can be used to estimate the limiting
value of $A_1^n(x)$ as $x  \rightarrow  1$, but perturbative QCD
alone does not predict a limiting value. Therefore, we prefer to
estimate the error due to the high-$x$ extrapolation by allowing
$|A_1^n(x)|  \le  1$, rather than by specifying a limiting value.
Thus we estimate a contribution $0.000\,\pm\, 0.003$
 from the $x>0.6$ region.
The low-$x$ extrapolation of the E142 data is $a$ $priori$ more
uncertain than that of the SMC, because the latter measure down to
lower $x$ = $0.006$. E142 have chosen to extrapolate using a power
form $g_1^n(x)$ = $Ax^{\alpha}$, where $\alpha$ was fixed at the
value $0.18$ predicted in ref.~\cite{Heimann} and adopted in
ref.~\cite{Schaefer}.
The model used in ref.~\cite{Schaefer} is however incompatible
with the new data, and hence cannot be used \cite{SchaeferI}
to support the value of $\alpha$ adopted by E142.
Other data do not fix the power $\alpha$: for example, a fit to
the EMC data at low $x$ gave \cite{EK}
$\alpha = 0.07^{+0.32}_{-0.42}$\,,
whereas we consider \cite{EK} a plausible
theoretical range to be $0 \le \alpha \le 0.5$, as assumed by the
SMC.
Using this latter range of $\alpha$ to estimate the contribution of the
range $x<0.03$ as $-0.006 \pm 0.006$ \footnote{We note that one
would obtain a lower value if one combined
E142 and SMC data, which are lower at small $x$.},
we estimate the full integral to be
\beq
\Gamma_1^n(Q^2{=}2\,\GeV)= -0.028
\,\pm\,0.006\,\hbox{(stat.)}\,\pm\,0.009\,\hbox{(syst.)}
\label{ourSLACG}
\eeq
to be compared with the E142 estimate of $-0.022 \pm0.011$.
\pr
The value of $\GIn$ at $Q^2=4.6$ \GeV\
can be extracted from the SMC data \cite{SMC}
together with the appropriate value of $\GIp$ in Table 1.
This results in
\beq
\Gamma_1^n(Q^2{=}4.6\,\GeV)= -0.076
\,\pm\,0.046\,\hbox{(stat.)}\,\pm\,0.037\,\hbox{(syst.)}
\label{ourSMCG}
\eeq
to be compared with the SMC estimate of $-0.08\,\pm0.04\,\pm\,0.04$.
\pr
There has been some question \cite{PRbound}
whether the SMC data at large $x$
are compatible with general positivity bounds derived within the
framework of the quark-parton model. An interesting bound is derived
in ref.~\cite{PRbound}
by eliminating the $u$ quark-parton contributions to
$g_1^p(x,Q^2)$ and $g_1^n(x,Q^2)$:
\beq
\left\vert 4\gIn - \gIp \right\vert =
\left\vert {15\over18} \Dd\xQ + {3\over18}\Ds\xQ \right\vert
\le
      {15\over18}\, d\xQ + {3\over18}\,s\xQ,
\label{PRboundeq}
\eeq
It should be noted that this bound could be violated by higher-twist
effects, that are expected to grow at smaller $Q^2$ and larger $x$.
An additional possible source of error is the assumption, made
in ref.~\cite{PRbound} that
$\gIp + \gIn = \gId$ locally in $x$.
The superficially similar
relation among the first moments used by the SMC \cite{SMC}
$\GIp + \GIn \simeq \GId/(1-1.5\omega_D)$, where $\omega_D=0.058$
is the probability of the deuteron to be in a $D$-state,
is well justified, but the local relation is violated
by smearing of the deuteron at large $x$.
\pr
We re-express the bound \eqref{PRboundeq}
in terms of the directly-measured quantities
$A_1^p(x)$ and $A_1^n(x)$ and the $Q^2$-dependent quantities $R(x,Q^2)$
and $\xi(x,Q^2)$ $\equiv$\break
  $F_2^n(x,Q^2)/F_2^p(x,Q^2)$:
\beq
\left\vert 4 \AIn \xi\xQ - \AIp \right\vert
\le  \left[ 1 + R\xQ\right]  \left[ 4 \xi\xQ -  1\right]
\label{ourPRbound}
\eeq
We have evaluated both sides of this version of the bound,
using the same fixed values $Q^2$ = $2$, $4.6$ \GeV\ on each side
to compare EMC data with the E142 and SMC data respectively.

The E142 data
satisfy the bound comfortably, whereas the SMC data are in
marginal disagreement, as shown in figure 2. We do not take seriously
the latter disagreement, in view of the large errors and the possible
contributions of higher twist effects. Shifting the central values
of the SMC data at $x  \ge  0.4$ so as to be consistent with the
bound \eqref{ourPRbound}
would not in any case change significantly our estimate
of $\Gamma_1^n(Q^2)$ based on the SMC data.
\pr
\section{Testing the Bjorken Sum Rule}
\pr
We are now in a position to evaluate the difference between
$\Gamma_1^p(Q^2)$ and $\Gamma_1^n(Q^2)$ at $Q^2$ = $2$ \GeV\ using
the EMC and E142 data, and at $Q^2$ = $4.6$ \GeV\ using the EMC
and SMC data:
\bea
\Gamma_1^{p-n}(Q^2{=}2.0\,\GeV)= 0.152\,\pm\,0.014\,\pm\,0.021\,
\nonumber\\
\label{evaluations}\\
\Gamma_1^{p-n}(Q^2{=}4.6\,\GeV)= 0.201\,\pm\,0.048\,\pm\,0.042\,
\nonumber
\eea
These are to be compared with the Bjorken sum rule, which takes the
following form when the leading-order perturbative QCD corrections
are included \cite{Kodaira}:
\beq
\GIpn \equiv
\int_0^1 d x \, \left[ \gIp - \gIn \right] =
{1\over 6} g_A\left[ 1-\alpha_s(Q^2)/\pi\right]
\label{pertBJsr}
\eeq
To estimate the appropriate values of $\alpha_s(Q^2)$ at $Q^2=2$,
4.6 \GeV, we take the latest estimate at $\alpha_s(m_{\tau}^2)=
0.330\ \pm\ 0.046$ \cite{alphasRef},
which is consistent with determinations at higher $Q^2$,
and use the leading order corrections to calculate
$\alpha_s(2,\ 4.6\,\GeV)=$ $0.371\pm0.058$, $0.304\pm0.039$,
together with $g_A=1.2573\pm0.0028$ \cite{RPP},
leading to the following values of the right-hand-side
of equation \eqref{pertBJsr}:
\bea
\Gamma_1^{p-n}(Q^2{=}2.0\,\GeV)= 0.185\,\pm\,0.004\nonumber\\
\label{rhsPertBJ}\\
\Gamma_1^{p-n}(Q^2{=}4.6\,\GeV)= 0.189\,\pm\,0.003\nonumber
\eea
However, these values cannot to be compared with the evaluations
\eqref{evaluations} before including subasymptotic ${\cal O}(1/Q^2)$
effects \cite{BBK} due to
mass corrections and higher-twist operators, which are particularly
important at the low average $Q^2$ of E142.
\pr
The mass corrections \cite{BBK}
are proportional to moments of $g_1^{p,n}(x)$,
which can be estimated using the data themselves:
\bea
\langle x^2 \rangle^{p} \equiv \int d x \,x^2 \gIp
\approx
\left\{\matrix{
0.0168\,,\qquad(Q^2=2.0\,\GeV)\cr
0.0130\,,\qquad(Q^2=4.6\,\GeV)\cr
}\right.
\nonumber\\
\label{x2moments}\\
|\langle x^2 \rangle^{n} | \equiv \left |\int d x \,x^2 \gIn \right |
< 7\times 10^{-5}.\nonumber\\
\nonumber
\eea
Errors in $\langle x^2 \rangle^{p}$ can be neglected, since
the mass corrections include an additional small factor
$(4/9) (m_N^2/Q^2)$ where $n_N$ is the nucleon mass.
Estimating the higher-twist corrections requires calculating the
reduced matrix elements
$\langle\langle U^{N S} \rangle\rangle$ and
$\langle\langle U^{S} \rangle\rangle$
of the local operators of spin one and
twist four \cite{HigherTwist},\cite{BBK}:

\bea
U_\mu^S &=&
             \bar u g \tilde G_{\mu\nu} \gamma^\nu u
           + \bar d g \tilde G_{\mu\nu} \gamma^\nu d
+{18\over 5}\,
\bar s g \tilde G_{\mu\nu} \gamma^\nu s\,,
\nonumber\\
\label{HTops}\\
U_\mu^{N S} &=&
  \bar u g \tilde G_{\mu\nu} \gamma^\nu u
- \bar d g \tilde G_{\mu\nu} \gamma^\nu d\,,
\nonumber
\eea

\beq
\langle N | U_\mu | N \rangle = s_\mu
\langle\langle U \rangle\rangle,
\qquad s_\mu \equiv \bar N \gamma_\mu \gamma_5 N\,.
\eeq
where $\tilde G_{\mu\nu} \equiv \epsilon_{\mu\nu\alpha\beta}
G_a^{\alpha\beta} {\lambda_a/2}$ and $\bar N$, $N$ are
nucleon spinors.
The matrix elements of the non-strange quark operators have been
estimated using sum rule techniques in \cite{BBK}: the possible
error quoted is estimated at
$50\%$ for $\Gamma_1^{p-n}$
and
$100\%$ for $\Gamma_1^{p+n}$,
and it has been suggested that the
matrix element of the strange quark operator should not be
larger than this uncertainty. Ref. \cite{BBK} found the following
magnitudes for the higher-twist correction:
\bea
\delta\Gamma_1^{p-n}(Q^2{=}2.0\,\GeV)=-0.022\,\pm\,0.011\nonumber\\
\delta\Gamma_1^{p-n}(Q^2{=}4.6\,\GeV)=-0.010\,\pm\,0.005\nonumber\\
\label{HTcorr}\\
\delta\Gamma_1^{p+n}(Q^2{=}2.0\,\GeV)=+0.017\,\pm\,0.017\nonumber\\
\delta\Gamma_1^{p+n}(Q^2{=}4.6\,\GeV)=+0.007\,\pm\,0.007\nonumber
\eea
The net contribution of the higher-twist and mass corrections to
$\GIp$ is quite small, becoming negligible at the $Q^2$
of the EMC experiment, and has the effect of decreasing the integral
slightly at smaller $Q^2$. This is consistent with the trend found
in Table~1
above on the hypothesis that $A_1^p$ is independent of $Q^2$,
though the inferred variation cannot be considered significant. The
net contribution of the $1/Q^2$ corrections to $\GIn$
are much larger, positive, and particularly significant for the low
$Q^2$ of E142. Thus it is not surprising that their value of $\GIn$
happens to be larger than that of the SMC. The net
correction to the Bjorken sum rule is important, exceeding $10\%$
at $Q^2=2$ \GeV:
\bea
\Gamma_1^{p-n}(Q^2{=}2.0\,\GeV)= 0.163\,\pm\,0.012\nonumber\\
\label{fullBJsr}\\
\Gamma_1^{p-n}(Q^2{=}4.6\,\GeV)= 0.180\,\pm\,0.006\nonumber
\eea
Comparing these theoretical estimates with the experimental
evaluations \eqref{evaluations},
we conclude that {\em the E142 and EMC data
are consistent with the Bjorken sum rule within less than one
standard deviation}.
This point is shown graphically in figure~3.
One way of phrasing this consistency is to quote the effective values of
$g_A$ extracted from the two combinations of experiments
by subtracting the subasymptotic corrections \eqref{HTcorr}
and removing the $\alpha_s(Q^2)$ correction appearing in
\eqref{pertBJsr}:
\begin{eqnarray}
\hbox{EMC \& SMC:}\qquad g_A^{\eff} = 1.39\,\pm\,0.38
\nonumber\\
\label{gAvalues}\\
\hbox{EMC \& E142:}\qquad g_A^{\eff} = 1.19\,\pm\,0.17
\nonumber\\
\end{eqnarray}
Combining all three experiments, we find
\beq
g_A^{\eff} = 1.22\, \pm \,0.15
\label{gAfit}
\eeq
{\em The Bjorken sum rule is consistent with all the available data
on polarized structure functions, and is now verified at the $12\%$
level.}
\pr
\section{Evaluation of the Quark Contributions to the Nucleon Spin}
\pr
Having reassured ourselves that the available data on polarized
nucleon structure functions are consistent with the Bjorken sum rule,
we now extract the contributions of the $u,d$ and $s$ quark flavours
to the total nucleon spin, and compare them with theoretical estimates.
We treat independently the EMC data on polarized protons, the SMC
data on polarized deuterons, and the E142 data on polarized $^3$He.
In the case of the SMC result, we fit the data after applying the
D-wave correction. In the case of the E142 result, we assume the same
model of $^3$He as in ref.~\cite{E142}
(see also ref.~\cite{Ciofi} for a detailed discussion of
$^3$He wavefunction in this context).
 Before fitting, we apply
to each value of the integral of $g_1(x,Q^2)$ the appropriate
higher-twist, mass and perturbative QCD corrections. We denote
the resulting corrected first moments
$\tilde\GIp$ and $\tilde\GIn$, which
correspond to the net charge-weighted sums of quark helicities.
 Thus we have
\bea
{1\over2}\left( {4\over9}\Du + {1\over9} \Dd + {1\over9}\Ds\right) =
\quad
\tilde \Gamma_1^{p}\quad\,\,\,\,\,\,\,
(Q^2{=}10.7\,\GeV)=+0.140\,\pm\,0.023\nonumber\\
{1\over2}\left( {1\over9}\Du + {4\over9} \Dd + {1\over9}\Ds\right) =
\quad
\tilde \Gamma_1^{n}\quad\,\,\,\,\,\,\,
(Q^2{=}\phantom{1}
2.0\,\GeV)=-0.056\,\pm\,0.015
\label{fixes}\\
{1\over2}\left( {5\over9}\Du + {5\over9} \Dd + {2\over9}\Ds\right) =
\tilde \Gamma_1^{p} +
\tilde \Gamma_1^{n}\quad
(Q^2{=}\phantom{1}
4.6\,\GeV)=+0.048\,\pm\,0.055\nonumber
\eea
We combine each of these inputs with \cite{RPP}
\beq
\Du - \Dd = g_A = 1.2573 \pm0.0028
\label{Dudvalue}
\eeq
and take \cite{FoverD}
\beq
F=0.46\,\pm\,0.01;\quad D=0.79\,\pm\,0.01;\quad
F/D = 0.58 \pm 0.02 \quad
\label{FDvalue}
\eeq
Each of eqs.~\eqref{fixes} and \eqref{FDvalue} can be rewritten as
linear constraint relating the values of $\Delta \Sigma$ and
$\Ds$.
In fig.~4 we plot the allowed regions in the
$\Delta \Sigma$ -- $\Delta s$ plane, corresponding to
these constraints.
The three data sets above yield the following compatible estimates
of the total quark contribution to the proton spin:
\begin{eqnarray}
\hbox{EMC ($Q^2=10.7$ \GeV):}
\qquad \Du + \Dd + \Ds = +0.16\,\pm\,0.21\nonumber\\
\hbox{SMC ($Q^2=\phantom{1}4.6$ \GeV):}
\qquad \Du + \Dd + \Ds = +0.06\,\pm\,0.25
\label{DudsValues}
\\
\hbox{E142 ($Q^2=\phantom{1}2.0$ \GeV):}
\qquad \Du + \Dd + \Ds = +0.29\,\pm\,0.14\nonumber
\end{eqnarray}
Ignoring any possible $Q^2$ dependence in $\Du+\Dd+\Ds$,
we combine these in quadrature to obtain
\beq
\Delta\Sigma \equiv
\Du + \Dd + \Ds
=0.22\,\pm\,0.10
\label{globalDuds}
\eeq
where the individual quark contributions are
\bea
\Du =+0.80\,\pm\,0.04 \nonumber\\
\Dd =-0.46\,\pm\,0.04 \label{DuDdDs}\\
\Ds =-0.13\,\pm\,0.04\nonumber
\eea
The global estimate of $\Delta s$ is approximately
three standard deviations
away from zero, providing good evidence that the \naive\ assumption of
ref. \cite{EJ}, namely that strange quarks should not contribute to
the nucleon spin, is false. This does not strike us as implausible,
since we have much more sophisticated models of nucleon structure
then when \cite{EJ} was written almost 20 years ago.
\pr
In particular, we note that the value \eqref{globalDuds}\
of $\Sigma\Delta q$ is
quite close to zero, which is the value expected in the class of
chiral soliton models pioneered by the computation \cite{BEK} in
the Skyrme model in the limit
of massless quarks and a large number of colours. It has been
suggested that corrections to this limiting value might amount to
about $30\%$ in the case of realistic quark masses and three colours
\footnote{See ref.~\cite{Ryzak} for discussion of $1/N_c$ corrections
and
ref.~\cite{SkyrmeNow} for a current assessment of the situation
in Skyrme type models.}
which would be consistent with the experimental value \eqref{globalDuds}.
We are not aware of any other dynamical model of nucleon structure which
{\em predicts} a value of $\Sigma\Delta q$ close to zero, though some
models are able to $accommodate$ it. For example, it has been
suggested \cite{Deltag}
that gluons inside the nucleon might be highly polarized,
and that this polarization would communicate itself to the quarks
via radiative corrections and the axial anomaly. Unfortunately,
this interesting suggestion does not give insight why $\Sigma\Delta q$
should vanish in any limit\footnote{For proposals in this
direction, see ref.~\cite{spinrefs}.}.
There have been some initial calculations of matrix elements
``from first principles"
using lattice techniques \cite{lattice0}-\cite{latticeII},
which are consistent within errors with our determinations.
\pr
It is also worthwhile noticing that $\Ds$ being nonzero is
a part of an interesting pattern \cite{IK}:
experiment indicates that certain strange-quark bilinear operators,
such as $\bar s \gamma_\mu\gamma_5 s$ have relatively large
matrix elements in the proton, while others are very small.
The presence of a substantial non-valence component of
$\bar{s} s$ pairs in the proton has some striking consequences.
One of these is the evasion of the OZI rule in the couplings of
$\bar{s} s$ mesons to baryons \cite{EGK}, leading to
surprisingly large branching ratios for $\phi$ production in
$\bar {p} p$ annihilation at rest \cite{pbarpI}.

\pr
\section{Interpretation}
\pr
We conclude that experiment and theory are converging on consistent
and quite quantitative decompositions of the spin of the nucleon.
The chiral soliton models pioneered by the Skyrme model are perhaps
unique in providing understanding of these results, which
disagree with \naive\ quark model ideas. The basic idea of these
chiral models is that a nucleon contains a very large number of
very light, relativistic quarks, that are best described by a
topological lump in an effective bosonic field. The data seem to
indicate that this is a better starting-point for understanding the
spin of the nucleon than the \naive\ picture of three non-relativistic
valence constituent quarks. However, the constituent quark model
does provide good understanding of many other aspects of nucleon
structure. We are therefore confronted with the challenge \cite{GM}
of relating constituent quarks to the chiral picture that works so
well for the polarized structure function data re-analyzed in this
paper. There is already some indication \cite{Kaplan}-\cite{staticQ}
that this program has good chances of success.

\bigskip
\begin{flushleft}
{\Large\bf Acknowledgements}
\end{flushleft}
We thank our experimental colleagues for enlightening discussions,
especially B.~Frois, V.~Hughes, J.~Lichtenstadt and S.~Rock.
We also thank V. Braun for comments regarding the estimate
of errors in higher-twist effects.
This research was supported in part
by grant No.~90-00342 from the United States-Israel
Binational Science Foundation (BSF), Jerusalem, Israel,
and by the Basic Research Foundation administered by the
Israel Academy of Sciences and Humanities.
\bigskip
\bigskip

\def\etal{{\em et al.}}
\def\PL{{\em Phys. Lett.\ }}
\def\NP{{\em Nucl. Phys.\ }}
\def\PR{{\em Phys. Rev.\ }}
\def\PRL{{\em Phys. Rev. Lett.\ }}

\newpage
\begin{flushleft}
{\Large \bf Figure Captions:}\\
\end{flushleft}
\vskip 0.5cm
Fig.~1.
(a)
The polarized structure function $\gIp$ at $Q^2=2.0$, 4.6 and
10.7 \GeV, obtained from the EMC $A_1^p$ data \cite{EMCNP},
smoothed and combined with the NMC \cite{NMC} data
for $F_2^p\xQ$ and SLAC parametrization \cite{RLT} of
$R\xQ\equiv \sigma_L/\sigma_T$.
Continuous curve: $Q^2=2$ \GeV, dots: $Q^2=4.6$ \GeV,
dot-dash: $Q^2=10.7$ \GeV.
    Compensation between the decrease at large $x$ and the increase at
small $x$ results in the relatively small $Q^2$-variation of $\GIp$
found in Table~1.
(b)
The polarized structure function $\gIn$ at $Q^2=2.0$, 4.6 and
10.7 \GeV, obtained from the E142 $A_1^n$ data \cite{E142},
smoothed and combined with the NMC \cite{NMC} data
for $F_2^d\xQ$ and $F_2^p\xQ$,
and SLAC parametrization \cite{RLT} of
$R\xQ\equiv \sigma_L/\sigma_T$.
Continuous curve: $Q^2=2$ \GeV, dots: $Q^2=4.6$ \GeV,
dot-dash: $Q^2=10.7$ \GeV.
Unlike in the case of $\gIp$, there is no
compensation between the large and small $x$, and this results
in the relatively large $Q^2$-variation of $\GIn$.

\hfill\break
\vskip 0.5cm
\noindent
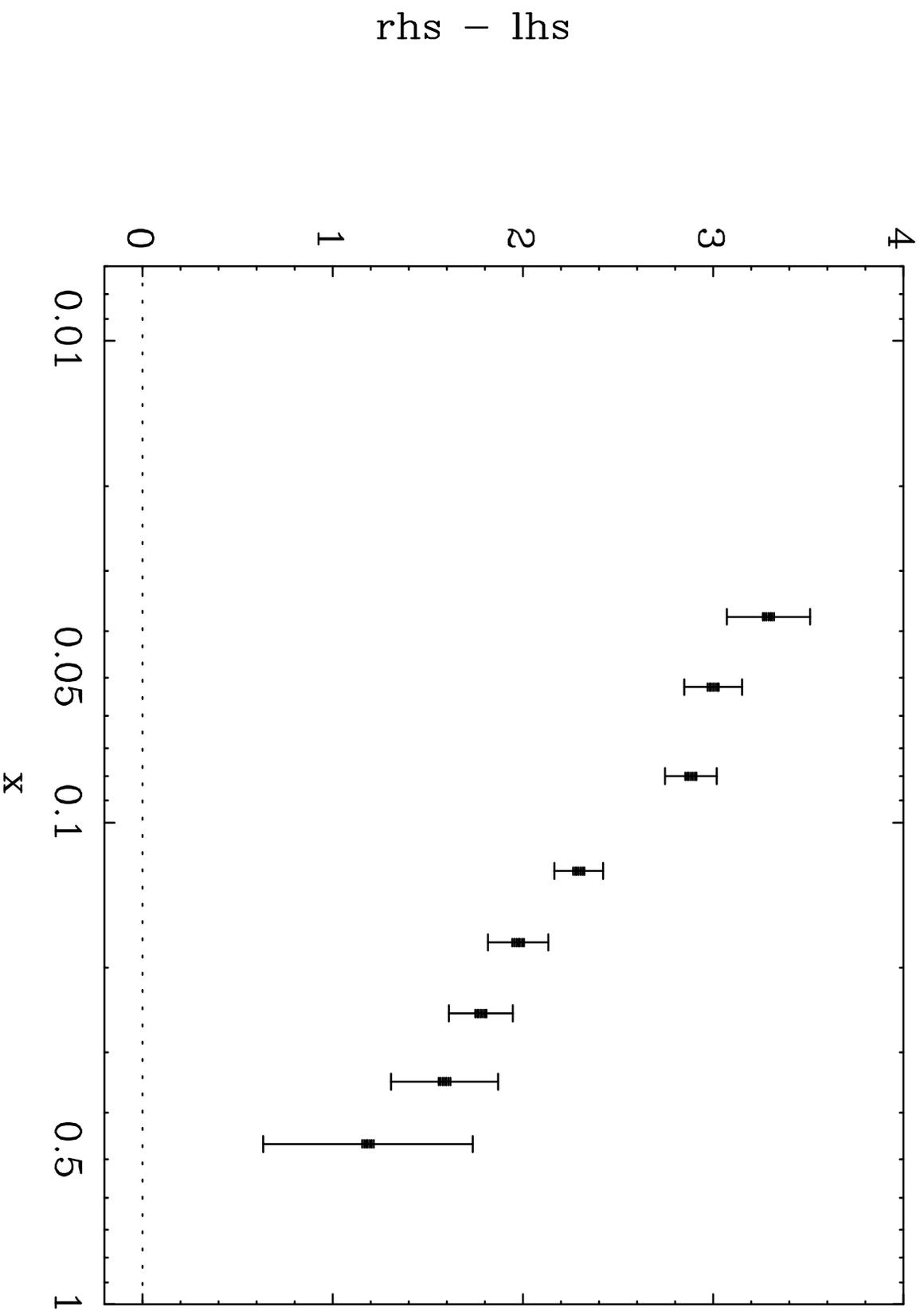
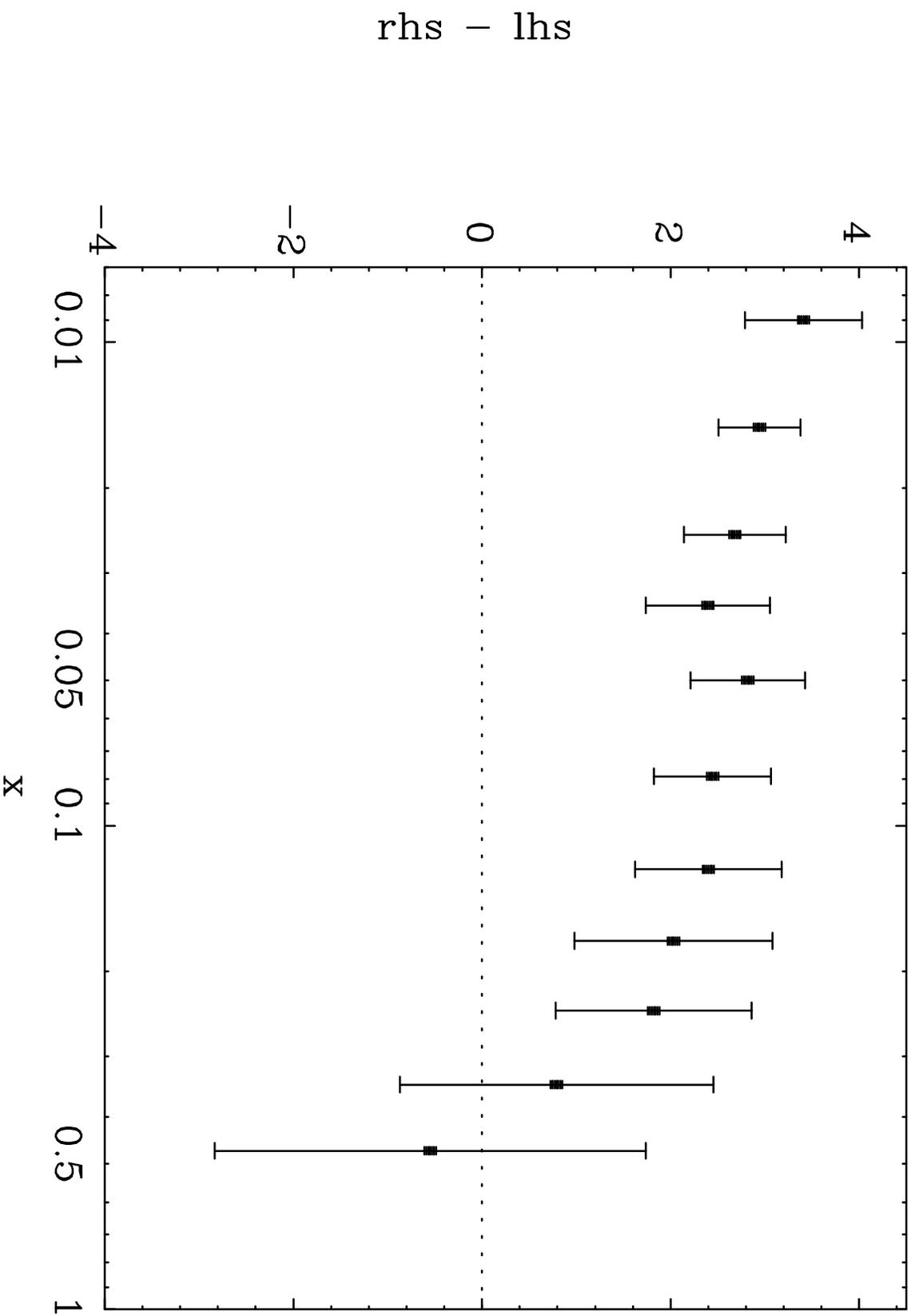
Fig.~2.
The difference between the right-hand and left-hand sides
of the bound \eqref{ourPRbound}. The actual errors
are slightly larger than those
indicated by the error bars, as the latter
refer to the error in the left-hand side only.
(a)~E142 data \cite{E142}, combined with EMC data \cite{EMCNP},
rescaled to $Q^2=2$ \GeV;
(b)~SMC data \cite{SMC}, combined with EMC data \cite{EMCNP},
rescaled to $Q^2=4.6$ \GeV.\hfill\break

\vskip 0.5cm
\noindent
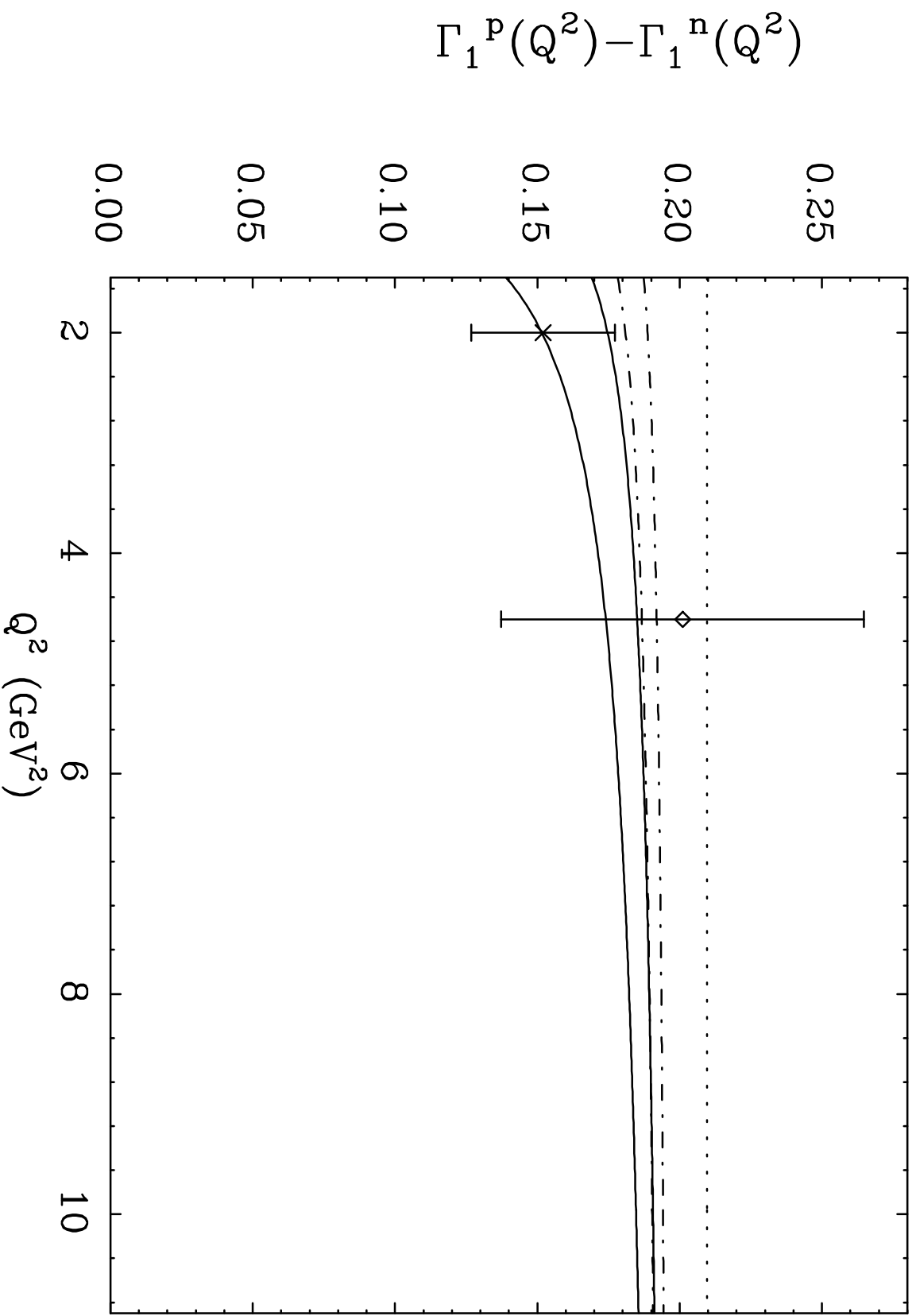
Fig.~3.
Experimental tests at $Q^2$ = 2 \GeV: E142 and EMC,
$Q^2$ = 4.6 \GeV: SMC and EMC of the Bjorken sum rule, including
perturbative QCD corrections (dot-dashed lines) and higher-twist
corrections (solid lines). The asymptotic value $g_A/6$ is denoted
by a dotted line.

\vskip 0.5cm
\noindent
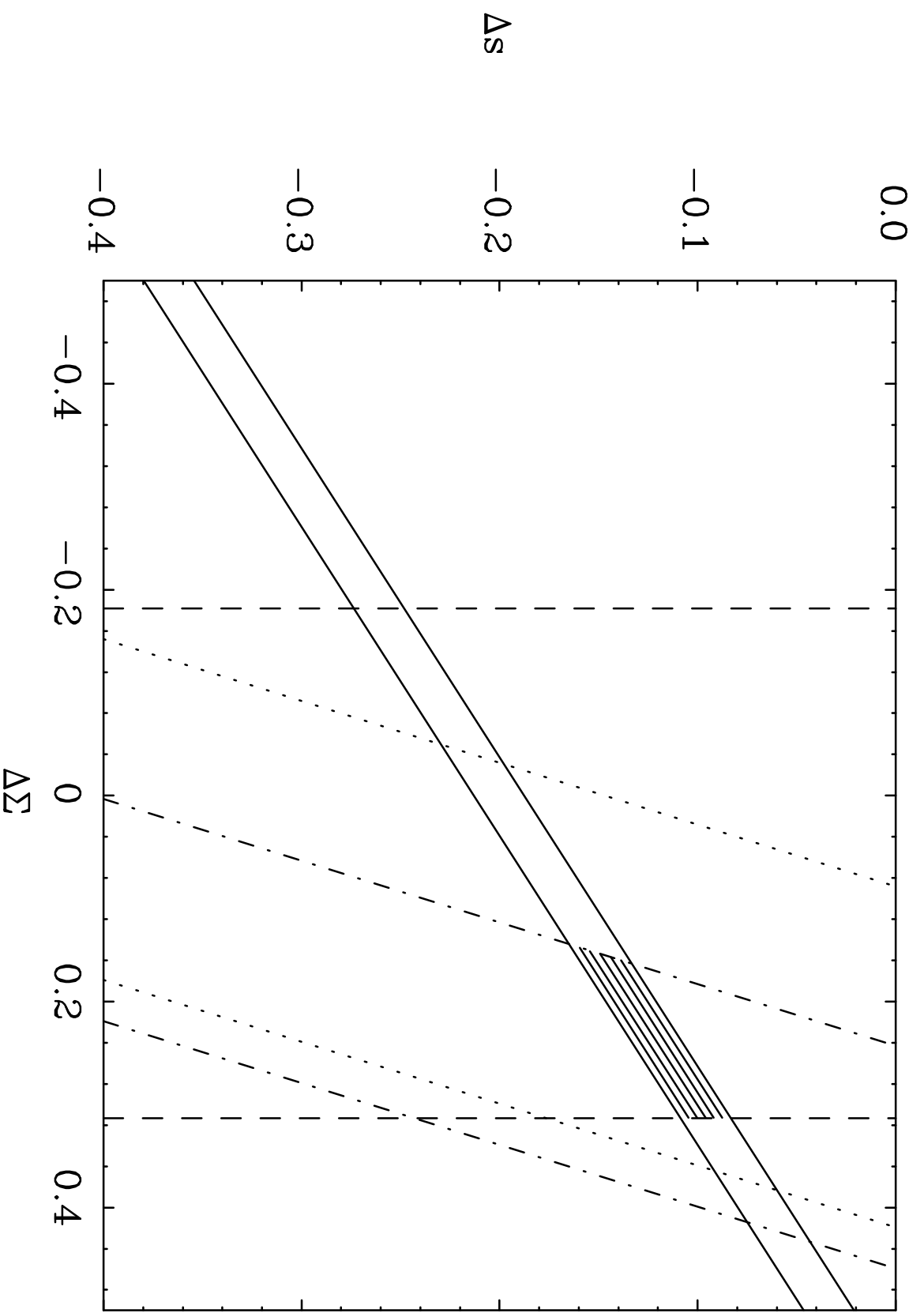
Fig.~4.
The allowed regions in the
$\Delta \Sigma$--$\Delta s$ plane, corresponding to
the linear constraints \eqref{fixes} and \eqref{FDvalue}.
Continuous lines: $\Delta \Sigma - 3 \Ds = 3F-D$;
dots: $\tilde \GIp$ constraint; dot-dash: $\tilde \GIn$;
dashes: $\tilde \GIp + \tilde \GIn$.

\end{document}